\documentclass[prl,superscriptaddress,lengthcheck,aps,amsmath,amssymb,preprintnumber,endfloats*]{revtex4}

\usepackage{graphicx}
\usepackage{dcolumn}
\usepackage{bm}
\usepackage{color}
\usepackage{psfrag}

\begin{document}
\preprint{\tt Not for circulation}

\title{Ground-State Fidelity and Bipartite Entanglement in the Bose-Hubbard model}

\author{P. Buonsante}
 \affiliation{Dipartimento di Fisica, Politecnico di Torino, Corso
 Duca degli Abruzzi 24, I-10129 Torino, Italy}%
\author{A. Vezzani}
 \affiliation{Dipartimento di Fisica, Universit\`a degli Studi di
 Parma, Parco Area delle Scienze 7/a, I-43100 Parma, Italy}
\date{\today}

\begin{abstract}
We analyze the  quantum phase transition in the
Bose-Hubbard model borrowing two tools  from quantum-information theory, i.e. the ground-state fidelity and  entanglement measures. 
We consider \mbox{systems} at unitary filling comprising up to 50
sites and show for the first time that a finite-size \mbox{scaling} analysis
of these quantities provides excellent estimates for the quantum critical point.
We conclude that fidelity is particularly suited for revealing a
quantum phase transition and pinning down the critical point thereof,
while the success of  entanglement measures depends
on the mechanisms governing the \mbox{transition}.
\end{abstract}
\maketitle

A few years ago some key works \cite{PACK1} initiated a new vein of research
using concepts borrowed from Quantum Information Theory in the analysis of  
quantum phase transitions (QPT,  i.e. phase transition driven
by quantum as opposed to thermal fluctuations).
The best known examples are no-doubt the  measures of entanglement, which quantifies the strength of quantum correlations between subsystems of a compound system
and represents a basic quantum-computational resource \cite{DiVincenzo_Nature_404_246}.
A more recent proposal is based on the fidelity, 
a key parameter in the characterization of the performance of logical quantum gates \cite{Schumacher_PRA_54_2614}.
The main advantage of this tool  lies in the fact that, being a purely Hilbert-space geometrical
quantity, it does not require any {\it a priori} knowledge of the 
correlations driving the QPT, or of the order parameter thereof  \cite{PACK0}.

While most of the works in this relatively new field focus either on fermionic models \cite{Gu_PRL_93_086402,Anfossi_PRL_95_056402,Anfossi_PRB_73_085113,Zanardi_QP_0606130,Cozzini_QP_0608059} or
on spin models \cite{PACK1,PACK0,PACK2,Vidal_PRA_69_054101,Vidal_PRA_69_022107,CamposVenuti_PRA_73_010303} that can be often effectively posed as free spinless fermionic systems, bosonic models went somewhat unaddressed so far.  Two exceptions in this respect are  Refs. \cite{Giorda_EPL_68_163} and \cite{Cucchietti_QP_0609202}, which propose the study of the hallmark QPT of the Bose-Hubbard (BH) model using respectively  entanglement and  Loschmidt Echo, a quantity kindred to fidelity. We note that the latter also provides an experimental scheme to measure fidelity. Also,  the crossovers characterizing the ground-state properties of the attractive BH model  \cite{Buonsante_PRA_72_043620} are investigated in terms of fidelity in Ref. \cite{Oelkers_CM_0611510}.
This substantial lack of attention does not make justice of this paradigmatic bosonic model. Indeed, on the one hand the BH model has a clear experimental relevance, being standardly realized in terms of optically trapped ultracold atoms \cite{Greiner_Nature_415_39}. On the other hand, it is a genuinely many-body model which in general cannot be reduced to an effective noninteracting model, hence posing a significant computational challenge. These features make the BH model ideal grounds for investigating the effectiveness of Quantum Information tools in the study of QPT.

This is the aim of the present work. We  focus two specific topologies, i.e. the one dimensional lattice with periodic boundary conditions (ring) and the completely connected graph (CCG), i.e. a model with the same hopping amplitude across any two sites. 
A twofold reason makes the latter  a convenient benchmark \cite{Lipkin_NP_62_188,Vidal_PRA_69_022107,Vidal_PRA_69_054101,Giorda_EPL_68_163,Zanardi_QP_0606130,Cozzini_QP_0608059}. First,  the critical lines of its zero-temperature phase diagram are known analytically in the thermodynamic limit \cite{Fisher_PRB_40_546}; second, its high degree of symmetry allows for a significant reduction of the relevant Hilbert space. 
As to the ring, a very interesting recent proposal \cite{Amico_PRL_95_063201} turned it from a convenient theoretical idealization to an experimentally 
realistic system \cite{Oelkers_CM_0611510,Buonsante_PRA_72_043620,Rey_CM_0611332}.
More specifically, we  investigate the relation between the behaviour
of the ground-state fidelity and bipartite mode entanglement --- both in the direct and reciprocal space --- and the superfluid-insulator QPT taking place in the {\it pure} BH model at integer filling. An efficient use of the system symmetries \cite{N:prep} allows us to apply exact diagonalization algorithms to rings and CCG's comprising up to 12 and 50 sites, respectively, and containing an equal number of bosons. We also consider the Gutzwiller mean-field approximation to the BH model, whose phase-diagram is essentially the same as that of the CCG \cite{Fisher_PRB_40_546}.
Our conclusions can be summarized as follows. The finite-size scaling of the position of suitably chosen extrema of both the fidelity and the bipartite entanglement  provides a satisfactory estimate of the critical point of the QPT, for both geometries. Consistent with what argued in the seminal Ref. \cite{PACK0}, fidelity turns out to be the candidate of choice when it comes to singling-out and quantitatively pinpointing the critical parameter of a possibly unknown QPT. Some of the appealing features that tip the scales in favour of fidelity in this task include: a very intuitive definition; a straightforward evaluation; a clear telltale of the transition; the robustness against mean-field approximation (note indeed that, unlike bipartite entanglement, the fidelity is defined also for the product trial state inherent in the Gutzwiller ansatz).
As to the bipartite mode-entanglement, its more complex definition 
and less straightforward evaluation make it a less handy tool for
 localizing the quantum critical points. Moreover, as it is
often pointed out in the literature (see e.g. Ref. \cite{CamposVenuti_PRA_73_010303})
a system-independent recipe based on entanglement measures seems to be lacking. However, the interest of such measures also relies in the insight that they can provide about the  mechanisms driving the (possibly otherwise identified) quantum phase transition \cite{Anfossi_PRL_95_056402}.
For instance, the  failure or success  of the recipes considered in this work can be explained based on the features of the QPT characterizing the system under examination.

The  Hamiltonian of the {\it pure} Bose-Hubbard model 
(i.e. on-site interactions, no offset in the local potential) reads
\begin{equation}
\label{E:BHH}
H = \frac{1}{2} \sum_{m=1}^M  n_m (n_m-1) - J \sum_{m,m'} a_m^\dag A_{m \,m'} a_{m'}
\end{equation}
where $a_m$, $a_m^\dag$ and $n_m = a_m^\dag a_m$ are lattice boson operators 
which destroy, create and count bosons at site $m$. As it was argued in Ref. \cite{Jaksch_PRL_81_3108} and later confirmed experimentally  \cite{Greiner_Nature_415_39}, Hamiltonian (\ref{E:BHH}) describes ultracold atoms trapped in an optical lattice and its  only effective parameter $J>0$, i.e.  the tunneling  amplitude to boson (repulsive) interaction ratio, is directly related to tunable experimental parameters such as the atomic scattering length and the intensity of the laser beams providing the optical confinement \cite{N:U}. The adjacency matrix $A$ describes the coordination of the $M$ sites composing the lattice, being nonzero only for adjacent sites. As we mention, we consider two topologies, i.e. the ring and the CCG, whose adjacency matrices read respectively $A_{m \, m'} = \delta_{|m - m'|,1}$, where $|M-1|=1$ owing to the periodic boundary conditions, and $A_{m \, m'} = 2 (1-\delta_{m,m'})/(M-1)$.
The normalization factor in the latter
 ensures that the generalized coordination number
$z_m = \sum_{m'=1}^M A_{m\,m'}$ equals 2 in both cases, so that in the thermodynamic limit of $M\to \infty$ the CCG is actually an {\it infinite range} mean-field approximation to the 1D system \cite{Fisher_PRB_40_546}. Indeed,
the critical boundaries in the phase diagram of the CCG \cite{Fisher_PRB_40_546}  are the same as those of the mean-field approximation ensuing from the decoupling assumption $a_m^\dag a_{m'} \approx a_m^\dag \langle a_{m'}\rangle + a_{m'}\langle a_{m}^\dag \rangle-\langle a_{m}^\dag \rangle\langle a_{m'} \rangle$, where $\langle \cdot \rangle$ denotes expectation on the ground state of the system \cite{Krauth_PRB_45_3137,Sheshadri_EPL_22_257}. For a homogeneous lattice like the ring, the resulting mean-field (or Gutzwiller) Hamiltonian is the sum of $M$ identical on-site contributions which, dropping the site subscripts, reads
\begin{equation}
\label{E:mfH}
{\cal H} = M \left[\frac{1}{2}  n (n-1) -\mu n- 2 J \alpha (a+a^\dag) \right]
\end{equation}
where $\alpha$ is the mean-field parameter that is related to the relevant ground-state $|\Psi\rangle$ by the self-consistency constraint $\alpha = \langle \Psi| a |\Psi \rangle$, and $\mu$ is the chemical potential, i.e. a Lagrange multiplier controlling the total number of bosons $N = \sum_m n_m$. Note indeed that since $[N,H]=0\neq [N,{\cal H}]$, the boson population is not a good quantum number in the mean field approximation. Conversely, the ground-state of Eq. (\ref{E:BHH}) can be studied within fixed-number Fock spaces. This reduces the in-principle infinite  Hilbert space of the system to the large but finite size ${\cal D}(M,N)= \binom{N+M-1}{N}$. The phase diagram of the BH model, usually drawn in the $\mu-J$ plane, comprises an extended compressible {\it superfluid} phase and a series of incompressible insulating {\it Mott lobes}. The latter correspond to commensurate populations,  i.e. to integer fillings $\nu=N/M$, and their
boundaries are given by the critical lines $\mu=\mu_{\pm}(J) = \pm [E_{\pm}(J) - E_{\rm c}(J)]$, where $E_{\rm c}$ and $E_{\pm}$ denote the ground-state energy
relevant to the commensurate ($N= \nu M$) and {\it defect} states ($N= \nu M \pm 1$), respectively \cite{PACK3,Elstner_PRB_59_12184}. On 1D system the transition at integer filling occurring at the {\it tip} of the lobe belongs to the particularly elusive Berezinskii-Kosterlitz-Thouless (BKT) class \cite{Fisher_PRB_40_546}. The precise location of the critical point thereof requires {\it ad-hoc} procedures that rely on an {\it a priori} knowledge of the mechanisms driving the transition \cite{PACK3}.
One of the best estimates for $\nu=1$, $J_\infty = 0.26 \pm 0.01$, has been obtained via numerical strong-coupling expansions of $\mu_\pm$ of remarkably high order \cite{Elstner_PRB_59_12184}. As we mention,   the boundaries of the mean-field Mott lobes are known analytically \cite{Fisher_PRB_40_546}. In particular, the tip of the $\nu=1$ Mott lobe is at $J_\infty = 3/2-\sqrt 2$. 

The critical point $J_\infty$ can be also studied considering  integer fillings only, $\nu = N/M \in {\mathbb N}$. Indeed, order parameters such as the superfluid or condensate fraction feature extrema in the suitable derivative with respect to $J$  \cite{Roth_PRA_68_023604} that are expected to diverge in the thermodynamic limit. A finite-size scaling of the location $J_M$ of such an extremum is expected to result in the critical point $J_\infty$,
\begin{equation}
\label{E:fit}
M = f(J_M) = C_1 |J_M -J_\infty|^{-\eta}, \qquad C_1,\eta > 0.
\end{equation}
 We checked this expectation and our results, which we do not show here, actually provide estimates of $J_\infty$ that agree very well with the best known estimate for the 1D lattice \cite{Elstner_PRB_59_12184} and the analytic result for the CCG. 

\begin{figure}[t!]
\begin{center}
\includegraphics[width=8.5cm]{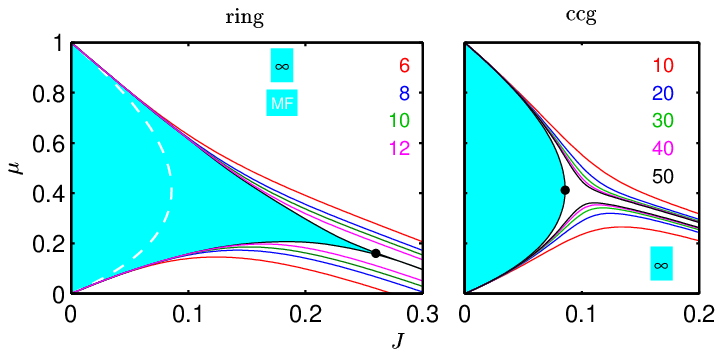}
\caption{\label{F:phdg} 
(color online) Mott lobe at filling $\nu=1$. Different colors correspond to different sizes as specified by the color code. The dot signals the critical point at filling $\nu=1$. Note that finite-size effects prevent the collapse of the critical boundaries.}
\end{center}
\end{figure}
As we mention, in this work we are interested in the insight provided
by observables borrowed from Quantum Information Theory, i.e. the
fidelity and the bipartite mode-entanglement. The former quantity has a remarkably simple definition, being nothing but the modulus of the overlap of two ground states relative to two different choices of the Hamiltonian parameters
${\cal F}(J,J') = |\langle\Psi_J|\Psi_{J'}\rangle|$.
On finite size systems, a drop in the fidelity corresponding to two arbitrarily
close parameter choices $J'=J+\delta J$ is expected to signal a precursor of the QPT \cite{PACK0,Zanardi_QP_0606130,Cozzini_QP_0608059}. A perhaps more effective indicator is provided by the peak in the ``density'' of the  second derivative of the fidelity ${\cal S}(J)= \lim_{\delta J \to 0} 2 [1-{\cal F}(J,J+\delta J)]/(M \delta J^2) $ \cite{Cozzini_QP_0608059}. The location $J_M$ of such peak on a size-$M$ system signals the QPT provided that the ${\cal S}(J_M)$ grows with increasing size, thus resulting in a divergence in the $M\to\infty$ limit.

The bipartite mode-entanglement (henceforth simply entanglement) is the von Neumann entropy
 ${\cal E}(J) = - {\rm Tr}
[\rho^*(J) \ln \rho^*(J)]/\ln{\cal D}^*$, where $\rho^*$ is the
reduced density matrix corresponding to one of the subsystems inherent
in a system bipartition, and ${\cal D}^*$ is the size of the relevant Hilbert space.
We  consider both the spatial modes (SM) $a_m$ and the quasimomentum
modes (QM) $b_q = M^{-\frac{1}{2}} \sum_m e^{i \frac{2 \pi}{M} m q}
a_m$. In both cases $\rho^*$ is obtained by tracing
out of the density matrix $\rho(J)=|\Psi_J\rangle\langle\Psi_J|$ the
degrees of freedom of all but one mode, so that ${\cal D}^*=N+1$. In the first case we choose one of the (equivalent) spatial bosonic modes. In the second case the untraced degrees of freedom correspond to the mode $q=0$, that describes the system ground-state in the noninteracting limit.

\begin{figure}[t!]
\begin{center}
\includegraphics[width=8.5cm]{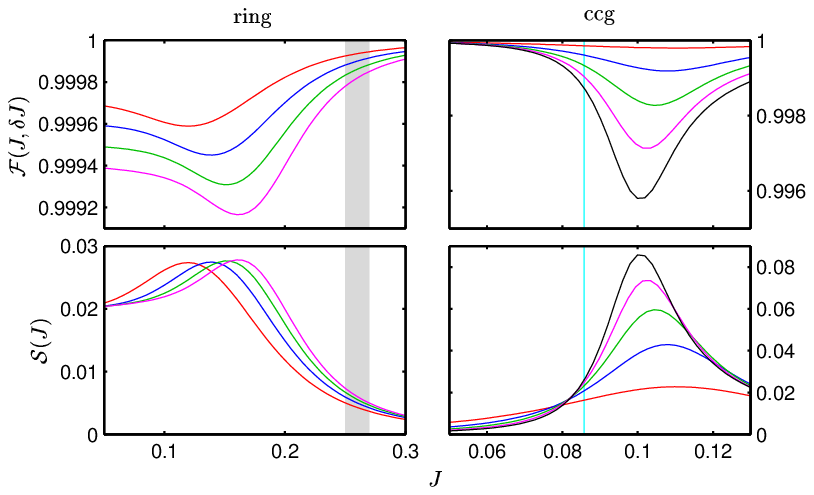}
\caption{\label{F:fid_wide}
(color online) Extrema of $\cal F$  and ${\cal S}$. The vertical stripe and line signal $J_\infty$ as reported in Refs. \cite{Elstner_PRB_59_12184,Fisher_PRB_40_546,Sheshadri_EPL_22_257}.}
\end{center}
\end{figure}

\begin{figure}[t!]
\begin{center}
\includegraphics[width=8.5cm,clip=true]{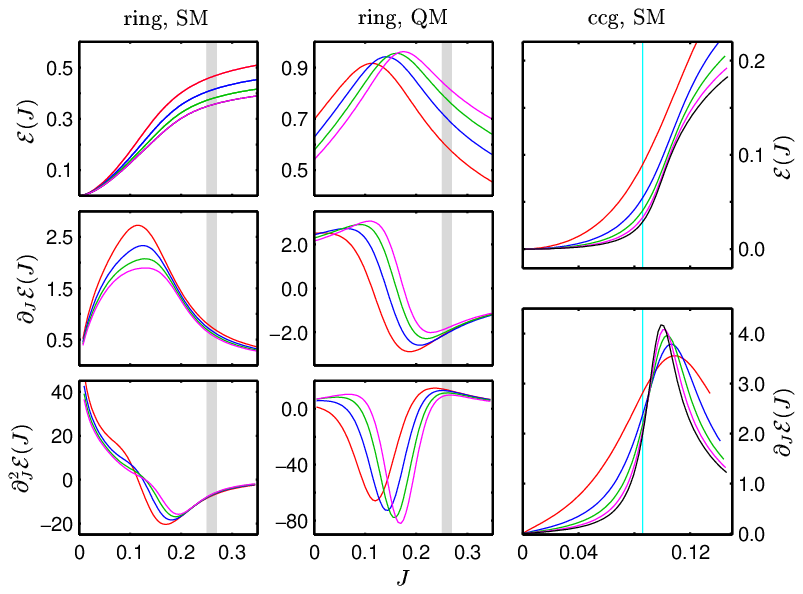}
\caption{\label{F:ent_wide} 
(color online). Extrema of $\cal E$ and its derivatives. 
The vertical stripes and lines signal $J_\infty$
as in Fig. \ref{F:fid_wide}.
}
\end{center}
\end{figure}

\begin{figure}[t!]
\begin{center}
\includegraphics[width=8.5cm]{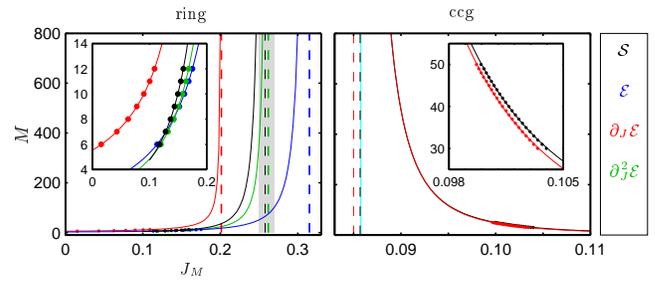}
\caption{\label{F:fits} 
(color online). Finite-size scaling of the extrema (dots)
in the considered quantities. Solid and dashed lines denote
the fits and the relevant estimates of $J_\infty$,
as specified by the color code. The entanglement in the ring (CCG) 
is evaluated with respect to QM (SM). The known values of $J_\infty$ are 
included as well.
}
\end{center}
\end{figure}

We now illustrate our results, which have been obtained by numerically
diagonalizing Eq. (\ref{E:BHH}). An efficient use of the Hamiltonian
symmetries allows us to consider rings (CCGs)  comprising up to $M=12$
($M=50$) sites at unitary filling. We remark that ${\cal
  D}(50,50)\approx 10^{29}$. Fig. \ref{F:phdg} shows the Mott lobe
corresponding to filling $\nu=1$ for both the ring 
and CCG.
 The filled regions represent the results in the thermodynamic
limit as reported in Refs. \cite{Elstner_PRB_59_12184} and
\cite{Fisher_PRB_40_546,Krauth_PRB_45_3137}, respectively. The solid
lines refer to finite-size results  as described by the color code,
which applies also to the subsequent figures. 
Fig. \ref{F:fid_wide} shows the behaviour of the fidelity ${\cal
  F}(J,J+\delta J)$  and of the relevant fidelity derivative density
${\cal S}(J)$ 
\cite{Zanardi_QP_0606130,Cozzini_QP_0608059,N:deltaJ} for the ring
 and CCG. The extrema of such quantities feature a
scaling behaviour that is consistent with the hypothesis that they
correspond to precursors of the quantum critical point. This is
signalled by a shaded stripe (corresponding to the estimate in
Ref. \cite{Elstner_PRB_59_12184}) and a vertical line for the ring and
the CCG, respectively. The same applies in Figs. \ref{F:ent_wide} and
\ref{F:fits}. Fitting the locations $J_M$ of these extrema as in
Eq. (\ref{E:fit}) results into $J_\infty = 0.257 \pm 0.001 $ ($J_\infty = 0.086
\pm 0.005$) for the ring (CCG) which agrees very satisfactorily with
the known result $J_\infty = 0.26 \pm 0.01 $
\cite{Elstner_PRB_59_12184} ($J_\infty \approx 0.0858$). This is
illustrated by the black plots in Fig. \ref{F:fits}. We also checked
the {\it extensivity} of the peak of the fidelity derivative, 
measured by the exponent $\gamma$ in the fit ${\cal S}(J_M) \sim C_2 M^\gamma$.
This is in general expected to be related with universal quantities and,
for the previously considered (effectively free) models, it  equals $1$
\cite{PACK0,Cozzini_CM_0611727,Cozzini_QP_0608059}.
 In the case of the ring we find a slightly but
definitely superextensive ($\gamma > 0$) behaviour, 
$\gamma = 0.087 \pm 0.009 $. 
In the case of the CCG, $\gamma = 0.749\pm 0.007$
signals a markedly superextensive behaviour. 

As it is clear from Fig. \ref{F:ent_wide}, entanglement is not an indicator
as clear as fidelity.
Indeed, considering SM modes
 of the ring, 
none of the entanglement derivatives seem
 to feature extrema displaying the expected scaling behaviour. 
The extrema, when present, become less pronounced with
 increasing size. The situation is more encouraging when the QM modes
 are considered.
Several extrema can be singled out
 that feature a scaling behaviour compatible with a QPT. However, as
 it is clear from Fig. \ref{F:fits}, a finite size scaling of these
 extrema according to Eq. (\ref{E:fit}) results in estimates of
 $J_\infty$  incompatible with each other. The fact that the critical
 point is known 
allows us to recognize that the
 minimum of $\partial_J^2 {\cal E}$ provides a satisfactory result,
 $J_\infty = 0.262 \pm 0.005$ (see also Fig. \ref{F:fits}). The
 rightmost column of Fig. \ref{F:ent_wide} shows that the situation
 for the CCG is considerably simpler. The maximum in the first
 derivative of the SM entanglement provides a satisfactory estimate for the critical point, $J_\infty = 0.085 \pm 0.007$. All of these features can be understood recalling 
 that the QPT at integer filling belongs to different universality classes for the ring and CCG. This is reflected also by the fact that in all of
the considered cases $J_M$ approaches $J_\infty$ from different sides depending on the connectivity of the lattice \cite{N:conn}.
In particular, the BKT  QPT characterizing the ring has been recently
shown to elude local measures of entanglement, such as the one based
 on SM \cite{CamposVenuti_PRA_73_010303}. In this respect, the global
 nature of the QM can be an explanation for the success of the relevant
   entanglement in capturing the transition. The
 less elusive nature of the generic mean-field QPT explains the
 success of SM in the case of CCG.

\begin{figure}[t!]
\begin{center}
\includegraphics[height=3.5cm]{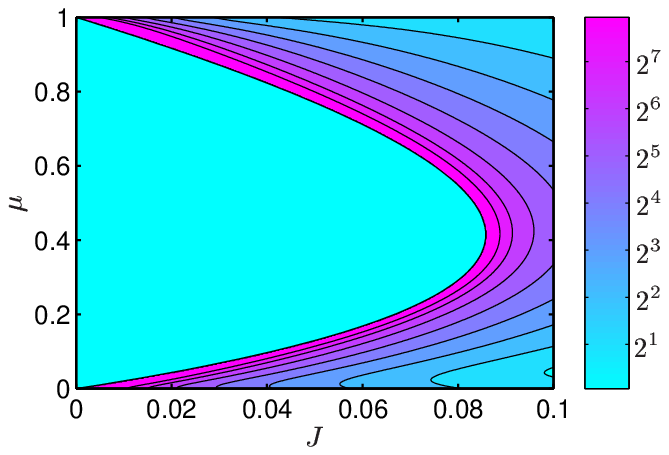}
\caption{\label{F:mf} Contour plot of ${\cal S}(\mu,J)$ for the ground-state of the mean-field Hamiltonian (\ref{E:mfH}).}
\end{center}
\end{figure}
The last result we present is the evaluation of ${\cal S}(J,\mu)$ for the ground state of the mean-field Hamiltonian (\ref{E:mfH}). This result is based on the (first order) analytical perturbative expansion of the ground-state with respect to $J$ and $\mu$ and the numerical evaluation of the mean-field parameter $\alpha$ \cite{N:prep}. The expected divergence of ${\cal S}$ at the superfluid side of the (mean-field) critical boundary is evident in the contour-plot in Fig. \ref{F:mf}.
 An analytic perturbative expansion in $\alpha$ in the proximity of such a boundary reveals that the exponent of the divergence is $-1$. Within the Mott lobe $\alpha=0$, so that  ground-state is independent of $J$ and $\mu$ and ${\cal S}=0$. Note that the mean-field approximation ``privileges'' a spatial representation, resulting in a factorized ground-state, which prevents the identification of the  QPT in terms of entanglement.

According to the above results, fidelity seems to be a very effective tool for investigating the presence of a QPT and pinning down the critical point thereof. Other than the intuitive definition and straightforward evaluation, the key asset of the fidelity approach is a clear telltale of the transition, namely a divergent behaviour of the maximum of ${\cal S}$ in the thermodynamic limit. Remarkably, this applies also for the elusive BKT transition of the ring. A further advantage of fidelity  is its applicability to approximated
versions of the model under examination, where entanglement measures are not viable.
Conversely, the more complex definition of the considered entanglement measure allows for a deeper insight in the nature of the transition and  underlying  quantum correlations \cite{Anfossi_PRL_95_056402}. A further remarkable result of this work is the precise estimate of $J_\infty$ for the  1D lattice.

The authors are grateful to  M. Cozzini for fruitful discussions and
 to C. Degli Esposti Boschi, M. Roncaglia and A. Anfossi for useful comments.
P. B. acknowledges a grant from the {\it Lagrange Project} --- CRT
 Foundation and useful comments and suggestions by  P. Giorda, P. Zanardi,
 I. Stojmenovich, V. Penna, P. Verrucchi and P.B. Blakie.


\begin{thebibliography}{10}

\bibitem{PACK1}
A.~Osterloh et~al., Nature {\bf 416}, 608 (2002); T.~J. Osborne and M.~A.
  Nielsen, Quantum Inf. Process. {\bf 1}, 45 (2002); Phys. Rev. A {\bf 66},
  032110 (2002); M.~C. Arnesen et~al., Phys. Rev. Lett. {\bf 87}, 017901
  (2001).

\bibitem{DiVincenzo_Nature_404_246}
D.~DiVincenzo and C.~Bennet,
\newblock Nature {\bf 404}, 247 (2000).

\bibitem{Schumacher_PRA_54_2614}
B.~Schumacher,
\newblock Phys. Rev. A {\bf 54}, 2614 (1996).

\bibitem{PACK0}
P.~Zanardi and N.~Paunkovic, Phys. Rev. E {\bf 74}, 031123 (2006); P.~Zanardi
  et~al., quant-ph/0701061 (2006).

\bibitem{Gu_PRL_93_086402}
S.-J. Gu et~al.,
\newblock Phys. Rev. Lett. {\bf 93}, 086402 (2004).

\bibitem{Anfossi_PRL_95_056402}
A.~Anfossi et~al.,
\newblock Phys. Rev. Lett. {\bf 95}, 056402 (2005).

\bibitem{Anfossi_PRB_73_085113}
A.~Anfossi et~al.,
\newblock Phys. Rev. B {\bf 73}, 085113 (2006).

\bibitem{Zanardi_QP_0606130}
P.~Zanardi et~al.,
\newblock quant-ph/0606130  (2006).

\bibitem{Cozzini_QP_0608059}
M.~Cozzini et~al.,
\newblock quant-ph/0608059  (2006).

\bibitem{Vidal_PRA_69_054101}
J.~Vidal et~al.,
\newblock Phys. Rev. A {\bf 69}, 054101 (2004).

\bibitem{Vidal_PRA_69_022107}
J.~Vidal et~al.,
\newblock Phys. Rev. A {\bf 69}, 022107 (2004).

\bibitem{CamposVenuti_PRA_73_010303}
L.~{Campos Venuti} et~al.,
\newblock Phys. Rev. A {\bf 73}, 010303R (2006).

\bibitem{PACK2}
G.~Vidal et~al., Phys. Rev. Lett. {\bf 90}, 227902 (2003); T.~Roscilde et~al.,
  {\it ibid}. {\bf 93}, 167203 (2004); {\bf 94}, 147208 (2005); S.~M. Giampaolo
  et~al., quant-ph/0604047 (2006).

\bibitem{Giorda_EPL_68_163}
P.~Giorda and P.~Zanardi,
\newblock Europhys. Lett. {\bf 68}, 163 (2004).

\bibitem{Cucchietti_QP_0609202}
F.~M. Cucchietti,
\newblock quant-ph/0609202  (2006).

\bibitem{Buonsante_PRA_72_043620}
P.~Buonsante et~al.,
\newblock Phys. Rev. A {\bf 72}, 043620 (2005).

\bibitem{Oelkers_CM_0611510}
N.~Oelkers and J.~Links,
\newblock cond-mat/0611510  (2006).

\bibitem{Greiner_Nature_415_39}
M.~Greiner et~al.,
\newblock Nature {\bf 415}, 39 (2002).

\bibitem{Lipkin_NP_62_188}
H.~J. Lipkin et~al.,
\newblock Nucl. Phys. {\bf 62}, 188 (1965).

\bibitem{Fisher_PRB_40_546}
M.~P.~A. Fisher et~al.,
\newblock Phys. Rev. B {\bf 40}, 546 (1989).

\bibitem{Amico_PRL_95_063201}
L.~Amico et~al.,
\newblock Phys. Rev. Lett. {\bf 95}, 063201 (2005).

\bibitem{Rey_CM_0611332}
A.~M. Rey et~al.,
\newblock cond-mat/0611332  (2006).

\bibitem{N:prep}
P. Buonsante et al., in preparation.

\bibitem{Jaksch_PRL_81_3108}
D.~Jaksch et~al.,
\newblock Phys. Rev. Lett. {\bf 81}, 3108 (1998).

\bibitem{N:U}
It is implicit in the form of Eq. (\ref{E:BHH}) that the boson-boson
  (repulsive) strength, usually denoted $U$, is our energy scale.

\bibitem{Krauth_PRB_45_3137}
W.~Krauth et~al.,
\newblock Phys. Rev. B {\bf 45}, 3137 (1992).

\bibitem{Sheshadri_EPL_22_257}
K.~Sheshadri et~al.,
\newblock Europhys. Lett. {\bf 22}, 257 (1993).

\bibitem{Elstner_PRB_59_12184}
N.~Elstner and H.~Monien,
\newblock Phys. Rev. B {\bf 59}, 12184 (1999).

\bibitem{PACK3}
G.~G. Batrouni et~al., Phys. Rev. Lett. {\bf 65}, 1765 (1990); T.~D.
  K{\"{u}}hner and H.~Monien, Phys. Rev. B {\bf 58}, R14741 (1998).

\bibitem{Roth_PRA_68_023604}
R.~Roth and K.~Burnett,
\newblock Phys. Rev. A {\bf 68}, 023604 (2003).

\bibitem{N:deltaJ}
In the left (right) plot we use $\delta J = 5\cdot 10^{-3}$ ($\delta J =2\cdot
  10^{-3}$).

\bibitem{Cozzini_CM_0611727}
M.~Cozzini et~al.,
\newblock cond-mat/0611727  (2006).

\bibitem{N:conn}
This interesting feature can be studied considering connectivities
  interpolating between the extremal values of the ring and CCG. The relation
  between $\gamma$ and universal quantities is also worth investigating.
  \cite{N:prep}.

\end{thebibliography}

\end{document}